# Femtosecond Coherence and Quantum Control of Single Molecules at Room Temperature

Richard Hildner,[1] Daan Brinks,[1] Niek F. van Hulst[1,2,*]

[1] ICFO – Institut de Ciencies Fotoniques, Mediterranean Technology Park, 08860 Castelldefels (Barcelona), Spain
[2] ICREA – Institucio Catalana de Recerca i Estudis Avancats, 08015 Barcelona, Spain

**Quantum mechanical phenomena, such as electronic coherence and entanglement, play a key role in achieving the unrivalled efficiencies of light-energy conversion in natural photosynthetic light-harvesting complexes, and triggered the growing interest in the possibility of organic quantum computing. Since biological systems are intrinsically heterogeneous, clear relations between structural and quantum-mechanical properties can only be obtained by investigating individual assemblies. However, single-molecule techniques to access ultrafast coherences at physiological conditions were not available so far. Here we show by employing femtosecond pulse-shaping techniques that quantum coherences in single organic molecules can be created, probed, and manipulated at ambient conditions even in highly disordered solid environments. We find broadly distributed coherence decay times for different individual molecules giving direct insight into the structural heterogeneity of the local surroundings. Most importantly, we induce Rabi-oscillations and control the coherent superposition state in a single molecule, thus performing a basic femtosecond single-qubit operation at room temperature.**

* Corresponding author:

  Phone:  +34 935534036
  Fax:    +34 935534000
  Email:  niek.vanhulst@icfo.es

Quantum coherence is the central aspect connecting fields as diverse as quantum computing and light-harvesting. The former relies on manipulations of coherent superposition states in 2-level systems to store and process information, surpassing the most powerful classical computers for several applications[1-4]. For light-harvesting nature exploits very long-lived electronic coherences of hundreds of femtoseconds in the initial steps of photosynthesis in bacteria and algae to achieve highly efficient and unidirectional energy flow towards reaction centres[5-7]. This observed longevity of coherences under physiological conditions questions the notion that interactions with the local environment universally lead to decoherence. A proper design of the surrounding may rather protect coherences, which opens interesting perspectives for natural quantum devices as the recently proposed concept of entanglement in light-harvesting complexes suggests[8]. Hence, unravelling the origins of coherence and decoherence on a nanoscopic scale is of high interest in many areas of physics.

Single-molecule detection combined with femtosecond pulse-shaping techniques provides a unique approach to gain insights into the ultrafast photophysics of individual quantum systems and to reveal correlations between e.g. structural properties and coherence times. Recently, we have succeeded in resolving incoherent vibrational relaxation[9] as well as in visualising excited state vibrational wave-packets[10] in single molecules at room temperature. Here we demonstrate that electronic coherences in single molecules can be established, probed, and controlled in disordered, non-crystalline environments under ambient conditions that are typical in e.g. biologically relevant systems. Our data reveal large variations in the time scales of the coherence decay for different individual molecules resulting from varying interactions with their particular local surroundings. In the coherent regime we are able to induce Rabi-oscillations, which was in solid-state systems so far only observed on individual molecules in crystalline matrices[11,12] and single quantum dots[13,14] at low temperatures. Moreover, we explore the limits of ultrafast manipulation of coherent superposition states by controlling the Bloch vector of a single molecule. This fundamental step constitutes a single-qubit operation at femtosecond time scales.

While at room temperature a very small number of molecules can be detected by stimulated emission[15], individual molecules can only be observed by their incoherent fluorescence signal as yet. The challenge in our experiments is therefore to translate information about variations of electronic coherences into changes of the spontaneous emission, i.e. of the excited state population probability. The basic idea of our approach is sketched in Fig. 1: A single molecule is resonantly excited into the purely electronic transition between the electronic ground ($|2\rangle$) and lowest excited state ($|1\rangle$) by femtosecond phase-locked double-pulse sequences. This induces stimulated absorption and emission processes and prepares this 2-level system in a certain state that is visualised by a vector (Bloch vector) on a sphere with unity radius (Bloch sphere, Fig. 1a)[16-18]. A vector pointing to the poles represents an eigenstate of the 2-level system, while any other position indicates a coherent superposition between levels $|1\rangle$ and $|2\rangle$. Interaction with the first pulse generally creates a coherent superposition state corresponding to a rotation of the Bloch-vector away from its ground state position ("south" pole, Fig. 1b). Depending on the experimental conditions several situations can then be distinguished. In the simplest case the molecule is completely isolated without interactions with its environment (Fig. 1b, left). If the phase difference $\Delta\phi$ between the pulses is fixed at 0 rad, the second pulse rotates the Bloch vector further about the same axis as the first pulse[17]. Since interaction with the light field is the only process changing the state of the Bloch vector, its final position is independent of the delay time $\Delta t$ (Fig. 1b, top left sphere). However, if the molecule is embedded in a disordered environment at room temperature (Fig. 1b, right), the phase memory between the ground and excited state wavefunctions is broken rapidly by pure dephasing processes caused by interactions with the matrix[19]. Here, the magnitude of the Bloch vector is not preserved and decreases with time. Consequently, the position of the tip after the pulse sequence changes, and importantly, the excited state population becomes smaller as a function of $\Delta t$ (Fig. 1b, top right sphere, green and blue arrows). The resulting decay of the spontaneous emission with increasing $\Delta t$ directly provides information about the coherent state of the molecule provided that population decay due to (non-) radiative transitions is negligible on the time scales of $\Delta t$. Key for the detection of coherence with this approach is the control of the relative phase $\Delta\phi$ between the electric fields of the laser pulses, since $\Delta\phi$ determines the rotation direction of the Bloch vector induced by the second pulse. Specifically, a $\pi$ phase shift inverts the rotation direction and moves the Bloch vector back towards its ground state position (Fig. 1b bottom, green arrows: $\Delta\phi = 0$ rad, black arrows: $\Delta\phi = \pi$ rad). This control of $\Delta\phi$ therefore offers the possibility to manipulate the motion of the Bloch vector before loss of phase memory and gives direct insight into the dephasing dynamics of single molecules.

As a model system we investigated single terrylenediimide (TDI) molecules dispersed in a film of poly(methyl-metacrylate), PMMA. Upon (near-)resonant excitation of the purely electronic transition of TDI with phase-locked double-pulse sequences (pulse width 75 fs, see Methods section) different molecules show distinct emission responses as a function of the delay time and relative phase. Examples of delay traces with $\Delta\phi$ fixed at 0 rad are depicted in Fig. 2a-d (black curves), with the time-averaged excitation intensity being kept constant during the acquisition of each trace. The emission signals in Figs. 2a,b,d feature pronounced variations of up to a factor of 2 with decay and recovery times, respectively, of several tens of fs, whereas the trace in Fig. 2c is constant within the signal-to-noise ratio. For $\Delta t > 300$ fs the emission of all investigated molecules remained constant.

The traces in Figs. 2a,b reflect the decay of coherences in single molecules with time constants of ca. 50 fs, because for TDI spontaneous emission as well as non-radiative transitions are negligible within the first 600 fs after excitation[20]. The turnover to flat and rising traces (Fig. 2c,d) is attributed to an increasingly stronger interaction between the laser field $\vec{E}$ and the transition dipole moment $\vec{\mu}_{12}$ between the electronic ground and excited state, i.e. an increasing Rabi-frequency $\omega_R = \dfrac{\vec{\mu}_{12} \cdot \vec{E}}{\hbar}$. These variations in $\omega_R$ come from different excitation intensities, random orientations of the TDI-molecules, and an intrinsic distribution of $\vec{\mu}_{12}$ caused by varying local surroundings. For a given $\Delta t$ this gives rise to different excited state probabilities from molecule to molecule, which in turn are reflected in the number of detected photons. In the examples in Fig. 2a-d the count rates increase from 400 to 2400 s$^{-1}$, and the highest observed rate was 4800 s$^{-1}$ (average between $\Delta t = 550$ and 600 fs).

These highest count rates are the maximum that can be recorded from a single molecule excited at saturation levels with a repetition rate of 500 kHz and a typical detection efficiency of 1 – 2 %.

To gain quantitative insights into the ultrafast dynamics we performed numerical simulations based on the optical Bloch equations in the rotating-wave approximation[17,18]. We calculated the elements of the density matrix $\rho$ for a 2-level system as a function of the delay time immediately after the pulse sequence interacted with the 2-level system. Pure electronic dephasing processes described by the time constant $T_2^*$ (relaxation of off-diagonal elements of $\rho$ or coherences) were included, while population decay (relaxation of diagonal elements of $\rho$ or populations) was neglected. In our situation the excited state population probability as a function of $\Delta t$, $\rho_{11}(\Delta t)$, is directly proportional to the fluorescence count rate in the recorded traces. To fit the data we varied the free parameters in the Bloch equations, i.e. the pure dephasing time $T_2^*$, the maximum Rabi-frequency $\omega_{R,0}$, and the detuning $\delta$, and minimised the residuals between the measured traces and the simulated $\rho_{11}(\Delta t)$-curves; see the methods section for details.

As evident from Fig. 2a-d the calculations (red curves) reproduce the data (black) very well and we find that $\omega_{R,0}$ increases from 0.01 to 0.06 fs$^{-1}$ (from top to bottom). The turnover from decaying to rising delay traces thus indeed results from an increasing Rabi-frequency. The decay (Fig. 2a,b) and recovery (Fig. 2d) time constants are determined by the pure dephasing time $T_2^*$ (convoluted with the finite pulse width), which is between 25 and 50 fs. The detuning $\delta$ is always 0 cm$^{-1}$ for the examples shown here. Importantly, also the flat trace in Fig. 2c can be unambiguously simulated. This curve cannot be modelled by allowing very long or short dephasing times ($T_2^* > 150$ fs or $< 20$ fs), because this would have to be compensated for by unreasonable $\omega_{R,0}$-values.

Since we compute the entire density matrix to fit the delay traces, the coherent state of each molecule can be reconstructed. The off-diagonal terms of $\rho$ [here: $i \cdot (\rho_{21} - \rho_{12})$] are depicted as a function of $\Delta t$ in Fig. 2e-h for the corresponding data in Fig. 2a-d. The coherences decay with increasing delay times and both the magnitudes and decay times are correlated to those of the emission signals. This demonstrates that the recorded fluorescence signals directly reflect the coherence decay in single molecules. We note that the coherences shown in Fig. 2e,f do not decay to zero for long delays, because for any given $\Delta t$ the density matrix elements were calculated immediately after the decay of the delayed pulse and the coherences prepared by this pulse did not completely vanish in these two examples.

The good agreement between data and theory justifies the approximations in the Bloch equations, and the femtosecond dynamics of single molecules is fully determined by $T_2^*$, $\omega_{R,0}$, and $\delta$. Using these parameters we can now directly visualise the time-dependent trajectories of the tip of the Bloch vector on the Bloch sphere[16,17] to gain further insight into the subtle differences in the interaction between the excitation pulse train and the molecules. The trajectories are displayed together with the corresponding traces in Fig. 2a-d for delays of $\Delta t = 0$ fs (green) and $\Delta t = 400$ fs (blue). In the trajectories in Fig. 2a,b the Rabi-frequencies are rather small and only absorption takes place. Here, the excited state population probability (z-component) after the pulse sequence decreases with $\Delta t$, since dephasing leads to less efficient interaction between the laser pulses and the increasingly shorter Bloch vector. Consequently, the observed emission signals decay with $\Delta t$. In Fig. 2c,d the substantially higher $\omega_{R,0}$ give rise to large nutation angles of the Bloch vector, and also stimulated emission during interaction with the laser pulses become important. Loss of phase memory renders this dumping of excited state population less efficient with increasing $\Delta t$, resulting in higher $\rho_{11}$–values at longer delays. The trajectory in Fig. 2d is particularly interesting because for $\Delta t = 0$ fs (green) the nutation angle of the Bloch vector is about $2\pi$. Thus the observation of a rising emission signal as a function of $\Delta t$ indicates that we monitor a full Rabi-cycle in a single molecule for $\Delta t = 0$ fs. This interpretation is verified by single-pulse experiments ($\Delta t = 0$ fs) on individual molecules, where we continuously increased the Rabi-frequency, i.e. the excitation power, and recorded the emission intensity. For several molecules we found first a rise and at higher excitation intensities a decrease of the fluorescence signal (see Supplementary Information, Fig. S1). This observation directly demonstrates Rabi-oscillations in single molecules at room temperature.

Having established coherent superposition states and monitored their evolution in single molecules at room temperature, the next step is to manipulate these states before the inevitable fast decoherence in disordered matrices. For delays shorter than $T_2^*$ a variation of $\Delta\phi$ between the excitation pulses changes the axis of rotation of the Bloch vector in a well defined way with respect to the rotation axis determined by the first pulse[17]. In particular, a π phase difference exactly reverses the sense of rotation (Fig. 1b, bottom). This provides a means to control the final coherent superposition state. In Fig. 3 a delay- and phase-dependent measurement (dots) on the same single molecule is displayed together with numerical simulations (lines), where the two pulses were in-phase (0 rad, green) and out-of-phase (π rad, black), respectively. The time-averaged excitation intensity was kept constant during the acquisition of both traces. Evidently, for long delays ($\Delta t \gg T_2^*$) the emission is independent of $\Delta\phi$ indicating a complete loss of phase memory, whereas before decoherence the count rate is strongly reduced by introducing the π phase shift. These data demonstrate that it is indeed feasible to manipulate the Bloch vector of single molecules despite fs dephasing times. In fact, pulse sequences with particular ($\Delta t, \Delta\phi$)-combinations ($\Delta t < T_2^*$) represent basic ultrafast single-qubit operations.

The fidelity of coherent state preparation and control is determined by the interplay between the pulse width and the pure dephasing times of single molecules. For 53 molecules we determined the dephasing times, which are between 25 and 110 fs with a peak at ca. 60 fs (see Supplementary Information, Fig. S2). This broad distribution reflects the heterogeneous nano-environments of single TDI-molecules in the amorphous PMMA-host. The dephasing times agree with values determined from the line width of room-temperature fluorescence spectra of individual chromophores of ca. 500 cm$^{-1}$ (FWHM), which constitutes an upper limit for the homogeneous line width and yields a lower boundary for $T_2^*$ of about 20 fs.

It is important to realise that in our experiments the dephasing time of each molecule is a time-averaged value, because we have to average over many excitation-dephasing-emission cycles to collect a sufficient number of photons for each ($\Delta t, \Delta\phi$)-combination. Individual "dephasing events" that destroy the coherent superposition state, i.e. (near-)elastic scattering with matrix vibrations[19], occur at a particular time after excitation. For consecutive excitation-dephasing cycles this time varies due to the disordered, temporally fluctuating surroundings of single molecules at ambient conditions. As a consequence, both the simulated density matrix elements as a function of $\Delta t$ as well as the time-dependent trajectories of the Bloch-vectors (Fig. 2) represent time-averaged curves. From the ergodic principle, stating that a time average is equivalent to an ensemble average, it follows that the coherence decays observed in our experiments (Figs. 2,3) can be identified with the envelope of the optical free-induction decay (OFID) of an individual molecule. The OFID is usually associated with the coherence decay in an ensemble of transition dipole moments[17,21,22]. In contrast to such bulk measurements, however, we are able to resolve variations in the decay times of the OFID for different individual molecules caused by varying local surroundings and interactions. These data constitute the first measurement of the OFID of a single quantum system in a solid, disordered matrix at room temperature. This was so far only observed on single nitrogen-vacancy electron spins in the crystalline and highly pure diamond environment[23].

The techniques introduced here can be applied to many types of single quantum systems, particularly natural photosynthetic light-harvesting complexes with their ten times longer electronic coherences[5-7]. For these multichromophoric systems coherent exciton dynamics and the influence of structural properties on electronic coherences can thus be probed on the level of single units even in highly complex and heterogeneous environments under physiological conditions. Together with the recent proposal of entanglement in such assemblies[8] this also opens fascinating routes towards ultrafast optical control of a single system representing a truly nanoscopic quantum device.

**Methods:**

*Experimental:*

Single terrylenediimide (TDI) molecules were dispersed in a poly(methyl-methacrylate) (PMMA) matrix at concentrations of ca. $10^{-10}$ M and investigated on a sample-scanning inverted confocal microscope. The excitation source was an optical parametric oscillator (OPO, Automatic PP, APE) that provided non-transform limited pulses with a width of 260 fs (spectral band width: 18 – 21 nm full width at half maximum, FWHM) at a repetition rate of 76 MHz. The centre wavelengths were between 622 and 640 nm to excite into (the high-energy tail of) the purely electronic transition of TDI (absorption max. in hexadecane: 643 nm[24]). A pulse picker (PulseSelect, APE) reduced the repetition rate to effectively 500 kHz (bunches of pulses with a repetition rate of 25 kHz, repetition rate within bunches: 4 MHz) to match it to the input of an acousto-optic programmable dispersive filter (Dazzler, Fastlite) that was used for pulse shaping. The shaper compressed the output of the OPO to transform-limited pulses with a width of 70 – 75 fs (FWHM) at the sample plane and generated pulse sequences with controlled delay time $\Delta t$ and phase difference $\Delta \phi$ between the electric fields of the output pulses. The shaped light was spatially filtered and collimated by a lens-pinhole-lens arrangement, directed into the microscope, and focussed on the sample by an oil-immersion objective with a high numerical aperture, NA = 1.3 (Fluar, Zeiss). To localise individual molecules a 10 x 10 $\mu m^2$ region of the sample was scanned, spatially well-separated molecules were selected, and then successively moved into the focus of the objective. The fluorescence light was split on two single photon counting avalanche photodiodes (APDs, Perkin-Elmer) by a polarising beam splitter to determine the degree of polarisation, i.e. the orientation of the transition dipole moment in the focal plane. The emission of each molecule was monitored as a function of $\Delta t$ and $\Delta \phi$ until photobleaching. For data analysis (see below) only molecules were considered, for which a trace could be measured at least twice. The excitation intensity at the sample was simultaneously recorded by a photodiode and was between 0.2 and 5 kW/cm$^2$. All experiments were carried out at room temperature under ambient conditions.

*Simulations:*

The data were simulated by the optical Bloch equations in the density matrix formalism employing the rotating-wave approximation (RWA). We neglected population relaxation (i.e. spontaneous emission and non-radiative decay) in the equations of motion for the diagonal terms of the density matrix $\rho$, since we are only interested in the dynamics up to 600 fs and TDI has an excited state lifetime $T_1$ of 3.5 ns in PMMA with a fluorescence quantum yield of $\sim 1$[20]. The excitation into (the high-energy tail of) the purely electronic transition allowed the energy level structure of TDI to be described by a simple 2-level system with the electronic ground ($|2\rangle$) and excited state ($|1\rangle$), Fig. 1. To account for this slightly off-resonant excitation as well as for the distribution of transition frequencies of individual TDI-molecules due to locally different interactions with the matrix[25], we included the detuning $\delta$. Owing to the pulsed excitation with similar values for the pulse width $\tau_p$ and the pure dephasing time $T_2^*$ we had to solve the Bloch equations numerically. The pulsed excitation is described by a time-dependent electric field amplitude $\vec{E}(t) = \vec{E}_0 \cdot f(t)$ with the peak value $\vec{E}_0$ and the envelope function $f(t) = \exp(-\frac{t^2}{2\tau_p^2}) + \exp(-\frac{(t-\Delta t)^2}{2\tau_p^2})$ of the double-pulse sequence in RWA. Consequently, the Rabi-frequency becomes time-dependent as well $\omega_R(t) = \frac{\vec{\mu}_{12} \cdot \vec{E}(t)}{\hbar}$, where $\vec{\mu}_{12}$ is the transition dipole moment. We define the time-independent maximum Rabi-frequency $\omega_{R,0} = \frac{\vec{\mu}_{12} \cdot \vec{E}_0}{\hbar}$, which is a free parameter in the simulations in addition to $T_2^*$ and $\delta$. The density matrix $\rho$ was evaluated as a function of $\Delta t$ (in steps of 5 – 10 fs) immediately after the decay of the envelope of the second delayed pulse ($\sim 3 \cdot \tau_p$ after the peak of the delayed pulse). Owing to the low repetition rate in our experiment both the excited state probability ($\rho_{11}$) and the coherences ($\rho_{12}$, $\rho_{21}$) completely decay before arrival of the next pulse pair. Therefore the initial conditions for the simulations at a given delay time are $\rho_{22}(0) = 1$ (only ground state occupied) and for all other elements $\rho_{ij}(0) = 0$.

We found that the density matrix as a function of the delay, $\rho(\Delta t)$, is very sensitive to the interplay between $\omega_{R,0}$, $T_2^*$, and $\delta$, which is a consequence of the comparable values of $\tau_p$, $T_2^*$, and the inverse maximum Rabi-frequency $1/\omega_{R,0}$ in our experiment (see Fig. 2). This sensitivity of the shape of the calculated traces to the fitting parameters is demonstrated by the comparison between the curves in Fig. 2c,d, which differ only in $T_2^*$. Hence, for the fits (Fig. 2a-d) the calculated $\rho_{11}(\Delta t)$-curves could be scaled such that their values at $\Delta t = 600$ fs match the average count rate of the experimental traces between $\Delta t = 550$ and 600 fs. Both curves can thus be overlaid without knowledge of the exact detection efficiency of our setup that depends, among other factors, on the detuning as well as on the unknown and random 3-dimensional orientation of the molecules. The best fit was determined by minimising the residuals between data and simulations. The coordinates of the Bloch vector tip used to calculate its time-dependent trajectories in Fig. 2 are related to the density matrix elements by

$x = \rho_{21} + \rho_{12}$, $y = i(\rho_{21} - \rho_{12})$, and $z = \rho_{11} - \rho_{22}$ [16,17].

**Acknowledgements:**

We thank T. H. Taminiau, F. D. Stefani, and F. Kulzer for discussions and assistance with the experimental setup, and K. Müllen for providing the molecules. Funding by the Körber foundation (Hamburg), the Spanish ministry of science and innovation (CSD2007-046-NanoLight.es and MAT2006-08184), and the European Union (FP6, Bio-Light-Touch) is gratefully acknowledged.

**Author Contributions:**

RH performed the measurements, analysed the data, and conducted the simulations. DB and RH constructed the experimental setup and performed control experiments. RH, DB, and NFvH conceived and designed the experiment, and discussed the data. RH wrote the paper. NFvH supervised the project.

**Figures:**

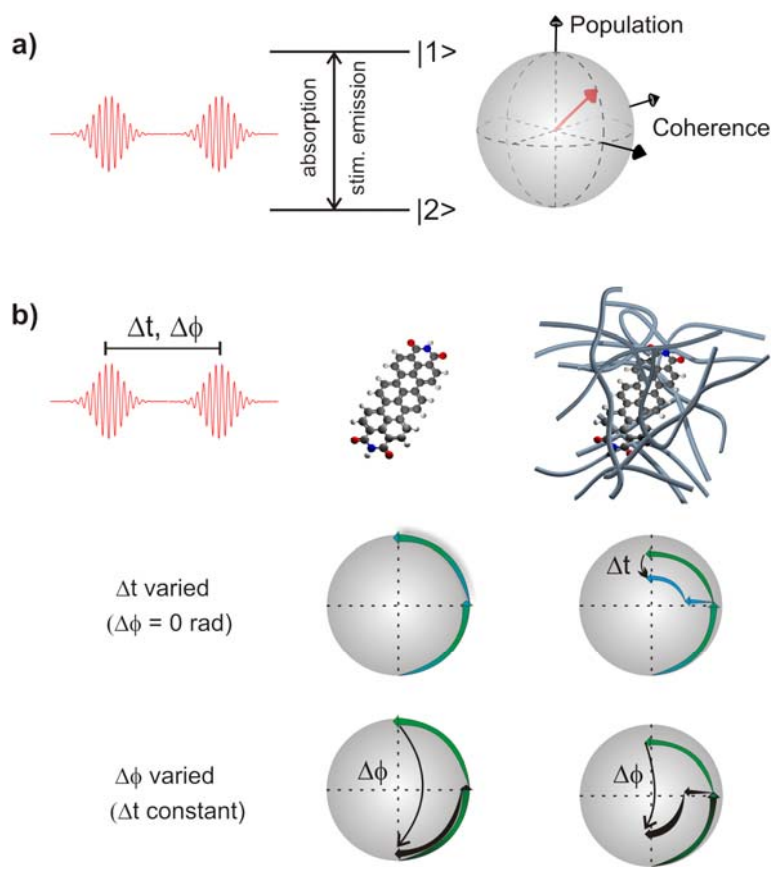

**Figure 1: Concept of the experiment. a)** A sequence of phase-locked ultrashort pulses resonantly drives a single molecule between the lowest excited (|1>) and the ground level (|2>). The state of this 2-level system is visualised by the Bloch-vector (red arrow) on the Bloch sphere, where the poles correspond to the eigenstates ("north": |1>, "south": |2>) and any other point indicates a coherent superposition between ground and excited state electronic wavefunctions. **b)** Influence of varying the delay time $\Delta t$ and relative phase $\Delta \phi$ on the trajectories of the tip of the Bloch vector: Without electronic dephasing a change of $\Delta t$ at constant $\Delta \phi$ does not affect the trajectory (top left sphere). In contrast, with dephasing the magnitude of the Bloch vector continuously decreases resulting in a measurable change in the exited state population for increasing $\Delta t$ (top right). The introduction of a phase change $\Delta \phi$ (at constant $\Delta t$) allows control of the coherent superposition state by altering the rotation direction of the Bloch vector (bottom). The fidelity of preparation and control of coherent superposition states is reduced with dephasing (right) as compared to the situation without dephasing (left).

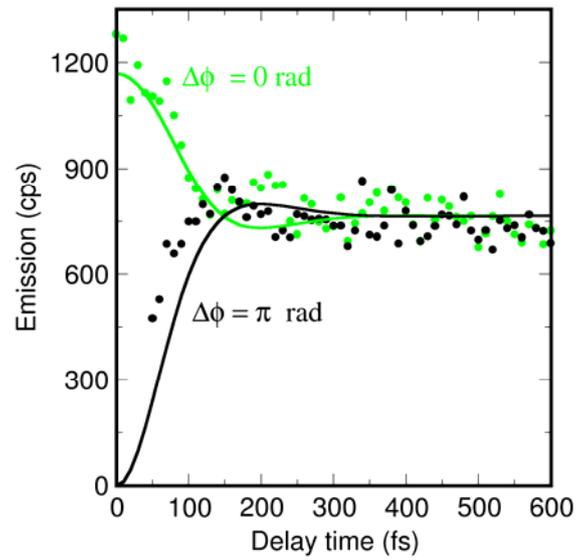

**Figure 3: Controlling the coherent superposition state of a single molecule.** Delay traces with $\Delta\phi$ = 0 rad (green) and $\Delta\phi = \pi$ rad (black) measured on the same molecule (cps: counts per second). For $\Delta\phi = \pi$ rad data points at $\Delta t < 50$ fs were not retrieved, because the excitation intensity could not be kept constant for these delays. The solid curves represent numerical simulations for the corresponding data yielding $\omega_{R,0} = 0.03$ fs$^{-1}$, $T_2^* = 60$ fs, and $\delta = 80$ cm$^{-1}$.

Supporting Information

# Femtosecond Coherence and Quantum Control of Single Molecules at Room Temperature

Richard Hildner,[1] Daan Brinks,[1] Niek F. van Hulst[1,2]

[1] ICFO – Institut de Ciencies Fotoniques, Mediterranean Technology Park, 08860 Castelldefels (Barcelona), Spain
[2] ICREA – Institucio Catalana de Recerca i Estudis Avancats, 08015 Barcelona, Spain

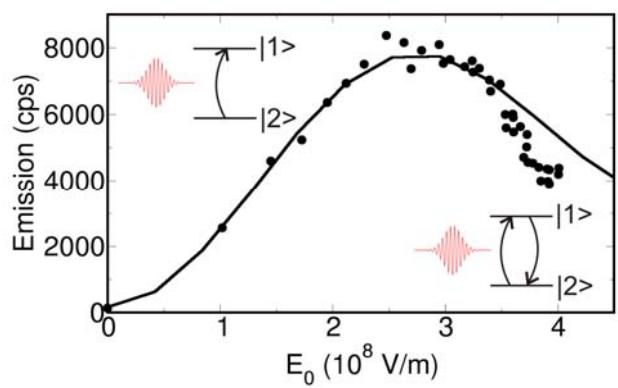

**Figure S1: Rabi-flopping of a single molecule.** Fluorescence intensity as a function of the peak electric field strength $E_0$ for a single molecule upon single pulse excitation ($\Delta t = 0$ fs, $\Delta\phi = 0$ rad, cps: counts per second). The solid curve represents the excited state probability from numerical simulations of the optical Bloch equations. The deviation between the simulation and the data at high peak field strengths comes from a slight thermally activated reorientation of the molecule, as observed by a small change of the degree of polarisation for these data points.

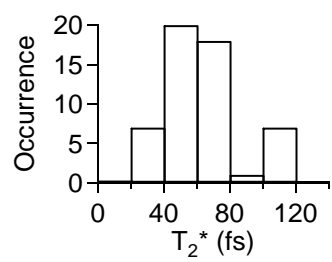

**Figure S2: Histogram of the pure dephasing times $T_2^*$.**